# Your Tweets Matter: How Social Media Sentiments Associate with COVID-19 Vaccination Rates in the US

Ana Aleksandric[1,2], Mercy Jesuloluwa Obasanya MPH[1], Sarah Melcher[1], Shirin Nilizadeh PhD[2], Gabriela Mustata Wilson PhD[1*]

[1] The University of Texas at Arlington, Multi-Interprofessional Center for Health Informatics, Arlington, TX, USA

[2] The University of Texas at Arlington, Department of Computer Science and Engineering, Arlington, TX, USA

## Abstract

**Objective:** The aims of the study were to examine the association between social media sentiments surrounding COVID-19 vaccination and the effects on vaccination rates in the United States (US), as well as other contributing factors to the COVID-19 vaccine hesitancy.

**Method:** The dataset used in this study consists of vaccine-related English tweets collected in real-time from January 4 - May 11, 2021, posted within the US, as well as health literacy (HL), social vulnerability index (SVI), and vaccination rates at the state level.

**Results:** The findings presented in this study demonstrate a significant correlation between the sentiments of the tweets and the vaccination rate in the US. The results also suggest a significant negative association between HL and SVI and that the state demographics correlate with both HL and SVI.

**Discussion:** Social media activity provides insights into public opinion about vaccinations and helps determine the required public health interventions to increase the vaccination rate in the US.

**Conclusion:** Health literacy, social vulnerability index and monitoring of social media sentiments need to be considered in public health interventions as part of vaccination campaigns.

**Abbreviations**: Health Literacy (HL), Social Vulnerability Index (SVI), Social Determinants of Health (SDOH), United States (US)
**Correspondence**: gabriela.wilson@uta.edu

## Introduction

Vaccines remain one of the most significant advancements and achievements in public health for disease prevention and control and one of the most cost-effective and successful interventions to improve health outcomes. Unfortunately, vaccine hesitancy from the public is a serious threat to maintaining herd immunity and preventing outbreaks [1]. The delay in acceptance or refusal of vaccines despite the availability of vaccine services is influenced by complacency, convenience, and confidence [2] and may be fueled by health information obtained from various sources, including social media [1]. A national survey in 2020 suggested that the public's willingness to vaccinate against COVID-19 was low and may be insufficient to provide herd immunity [3]. In the US, as of January 2021, about 20% of the population remained hesitant to get the vaccine, and 31% say they will wait to see how it is working for others before getting the COVID-19 vaccine [4].

Although social media is a valuable tool for disseminating and receiving relevant health information for patients, clinicians, and scientists [5], it has also negatively impacted vaccination rates and public health promotion. The substantial spread of negative posts across different social media platforms, such as Twitter and Facebook, on the vaccine's safety has fueled the public's hesitation in getting vaccinated [6]. A relevant example is a study by Ahmed *et al.* that demonstrated that using Twitter and Facebook as sources of information related to the influenza virus has a significant inverse association with influenza vaccine uptake [6]. The anti-vaxxer propaganda has also seen a rise in the community on social media and has potentially magnified the hesitancy [7].

Since the connection between social media and vaccine hesitancy has already been established in previous studies [6], it is hypothesized that a similar pattern occurs related to COVID-19 vaccine uptake. During the COVID-19 pandemic and the country's lock-down, people had to rely on social media to keep social interactions and connections going as they could not do so in person. At the same time, misinformation regarding COVID - 19 emerged in other regions around the world [8]. Unfortunately, this exposed the public to unsubstantial rumors regarding protective measures against the spread of the virus and COVID-19 vaccines. In particular, studies suggest that the use of social media as a source of information about COVID-19 has been correlated with stronger beliefs in conspiracy theories, adverse information, and less-protective behaviors during the pandemic [5]. Some of the false information being spread is that 5G mobile networks were connected to the spread of the virus, that vaccine trial participants have died after taking a COVID-19 vaccine, and that the pandemic is a conspiracy or a bioweapon [9]. As a result, a decreased confidence in the vaccine's efficacy and willingness to take the vaccine once available have been observed. Therefore, it is crucial to understand the impact of social media posts on vaccination rates, which can help identify intervention areas and address misinformation and disinformation. It is also essential to investigate the association between vaccination rates, health literacy, and social determinants of health, which refers to conditions in the places where people live, learn, work, and play that affect a wide range of health and quality-of life-risks and outcomes [10]. These factors could impact both health and health care disparities outcomes [11] and potentially influence vaccination rates in the United States.

While there have been studies/reports on social media and vaccine hesitancy [1,3,5,9], no study has examined the association between sentiments (positive and negative) and vaccination rates using primary data collected from a social media platform. This study addresses this gap by obtaining sentiments of vaccine-related tweets on Twitter and conducting a multivariant regression analysis to examine the impact on vaccination rates. The study also examines the association between social determinants of health factors – health literacy and social vulnerability index, with vaccination rates and analyzes how social media sentiments correlate with COVID-19 vaccine hesitancy and vaccination rates in the United States. It also contributes to a better understanding of the impact of social media and its potential role in healthcare communication and identifies opportunities for interventions such as addressing miscommunications and increasing health literacy to improve health. It also demonstrates the importance of incorporating social determinants of health factors (health literacy and social vulnerability index) in public health interventions in the future.

## Methods

### Data Sources
The various data sources used in this study are listed in **Table 1**.

**Table 1:** Data sources used in the analysis

| Data | Data Source |
|---|---|
| Tweets | CoVaxxy Dataset. http://arxiv.org/abs/2101.07694 |
| Health Literacy | The University of North Carolina at Chapel Hill. http://healthliteracymap.unc.edu/# |
| Social Vulnerability Index | CDC. https://www.atsdr.cdc.gov/placeandhealth/svi/index.html |
| Population per census block group | Policy Map. https://uta-policymap-com.ezproxy.uta.edu/newmaps#/ |
| Population per state | Policy Map. https://uta-policymap-com.ezproxy.uta.edu/newmaps#/ |
| Population Percentages | United States Census Bureau. https://www.census.gov/data/tables/time-series/demo/popest/2010s-state-detail.html |

### Study Design
This study focuses on one of the most popular microblogging and social networking service platforms, Twitter [12]. The dataset used herein comes from the CoVaxxy dataset [13], an extensive collection of vaccine-related tweets collected in real-time between January 4 - May 11, 2021, when the COVID-19 vaccine was approved and vaccination started. The primary goal of this research was to determine how tweets sentiments impacted the vaccination rates in the United States and determine what other factors, such as health literacy (HL) and social vulnerability index (SVI), might influence vaccination rates. The study was also performed to find the association between HL and SVI, as well as between HL, SVI, and state demographics (percentages of White, Black, Hispanic, and Asian populations per state).

### Study Population
The dataset was comprised of tweets from 51 individual states within the US, and from the collection of 85,100,935 vaccine-related tweets between January 4-May 11, 2021, 322,035 tweets were extracted with the specified location within the United States, belonging to 117,258 unique users. Twitter provides location as a bounding box of four points, each having longitude and latitude. However, for the purpose of this analysis, it was important to identify the state where the tweet was posted from. Therefore, the middle point of the bounding box (see **Figure 1***)* was selected, and the Federal Communication Commission getArea API [14] was used to determine the state of every tweet. Not all tweets include such information; therefore, the tweets that received 'None' as a location were discarded from the dataset.

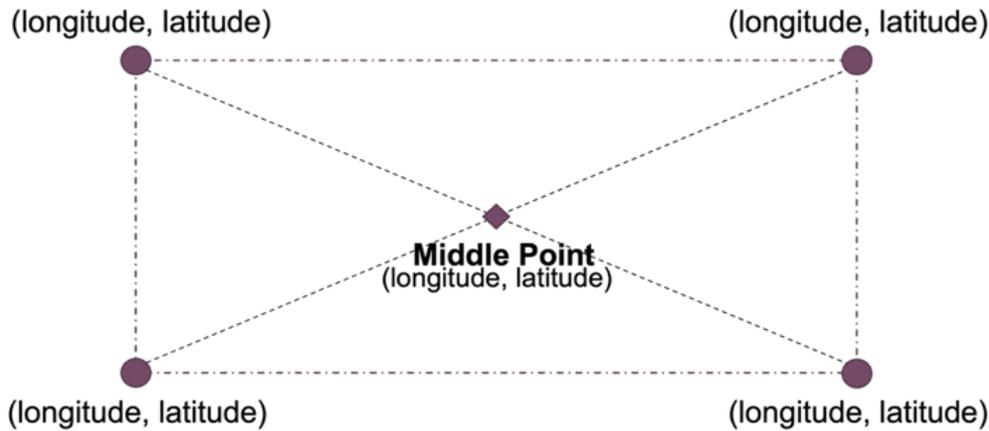

**Figure 1:** Tweet's bounding box location visualization; computing the middle point

One caveat is that bots can automatically generate tweets, which might not represent the opinion of people. Therefore, to remove any bias from the dataset, Botometer v4 [15] was used to remove any tweets belonging to the bot accounts. According to Botometer creators, between 9 and 15 percent of Twitter accounts are bot accounts, while 85 percent of the accounts are human [16]. Therefore, the $85^{th}$ percentile of scores was computed on all the accounts scores in the dataset being 0.69, which is then used as a threshold. Afterward, every tweet belonging to the account with a score higher than the threshold was removed from the dataset. This led to the removal of 58,263 tweets belonging to 17,597 users. Secondly, since the primary purpose of this research was to establish a connection between tweet sentiments and vaccination rates, all the tweets originating from the accounts representing organizations (e.g., pharmaceutical companies) were removed using Humanizr, a tool that can distinguish between personal and organizational Twitter accounts considering different account features [17]. The filtered dataset included the total number of 243,202 tweets belonging to 95,292 unique users. **Figure 2** highlights the process of data collection and the cleaning process.

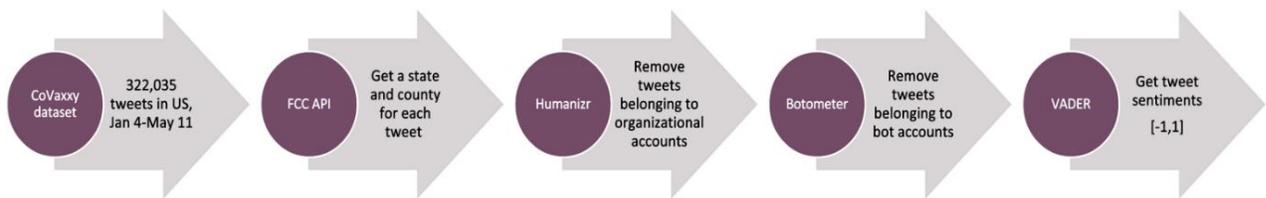

**Figure 2:** Data Collection and Cleaning Process

**Dependent and Independent Variables**
**Vaccination rate**: The main dependent variable in this analysis is the vaccination rate, as the objective is to see if tweets' sentiments and social determinants of health impact the vaccination rate in the United States. The vaccination rate shown in **Figure 3** is per hundred people on the last day of our data collection, May 11, 2021. The vaccination rate was extracted from COVID-19 vaccine data, collected daily in real-time by Indiana University's Observatory on Social Media (OSoMe) in collaboration with colleagues from Politecnico di Milano [13].

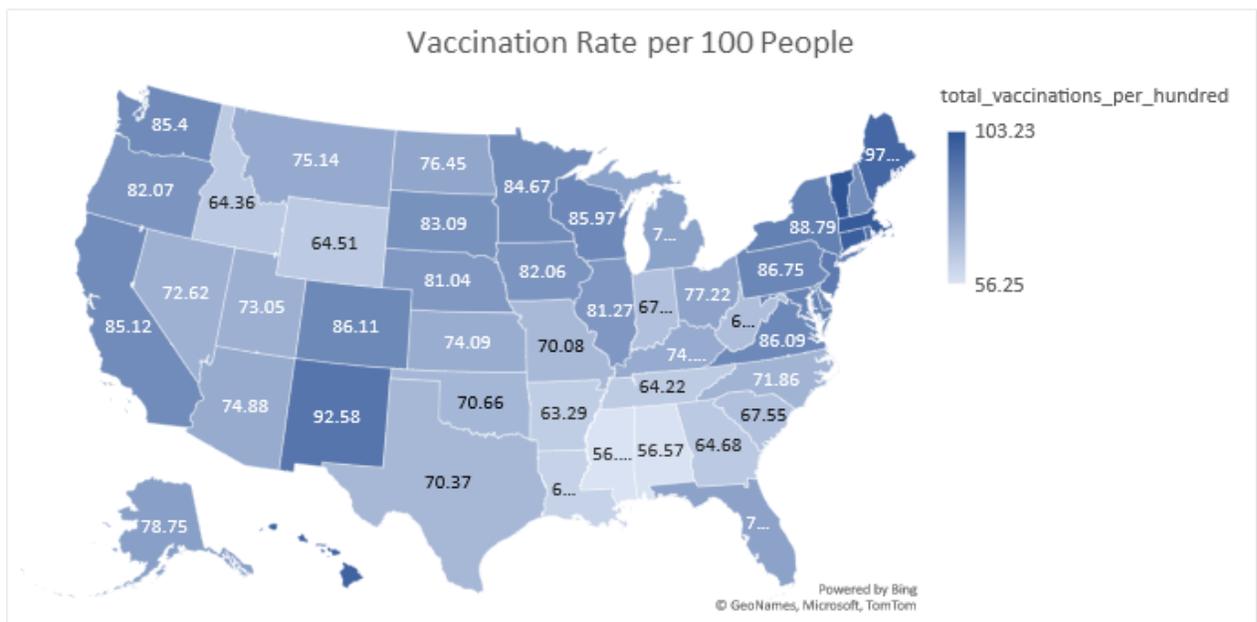

**Figure 3:** Vaccination rate in the United States per 100 people as of May 11, 2021 (figure generated with Microsoft Excel)

**Tweets' sentiments**: The sentiment of each tweet was extracted using VADER [18], a machine learning tool that allows for deriving sentiments of social media posts in multiple languages. VADER showed as the most accurate tool used for sentiment analysis on the Twitter data [19] due to its ability to take the social media language common features into consideration. This tool accepts a tweet text as an input, and it returns a compound score in the range from –1 to 1, where scores closer to 1 are more positive, and scores closer to -1 are more negative.

**Health literacy**: Health literacy is the degree to which individuals have the ability to find, understand, and use information and services to inform health-related decisions and actions for themselves and others [20]. Health literacy was used in the analysis to provide a better understanding of how the ability of the community to receive and understand health information is impacting the public health interventions. The assumption is that vaccine hesitancy is lower in the communities with high health literacy and higher in the communities with lower health literacy. Health literacy data was obtained from the University of North Carolina at Chapel Hill at the census block level [21]. A weighted average of health literacy indexes having census group population as weights was performed to calculate health literacy at the state level. Census group population was based on the 2010 census since health literacy data corresponds to 2014, which means health literacy data is calculated based on the 2010 census group data.

**Social vulnerability index**: The social vulnerability index refers to the potential negative effects on communities caused by external stresses on human health [22]. Such strains include natural or human-caused disasters or disease outbreaks. Low index means communities are especially at risk during public health emergencies because of socioeconomic status, household composition, minority status, or housing type and transportation [22]. The Center for Disease Control and Prevention (CDC) pointed out that areas with a high social vulnerability index had lower vaccination rates [23]. The Social Vulnerability Index (SVI) data used in this study originate from the CDC's SVI at the county level. Similarly, as in the case of HL, a weighted average of social vulnerability indexes was computed, having county population as weights to obtain SVI on the state level.

Given that population data were not available for the Oglala Lakota and Kusilvak, these counties were excluded from the calculation of weighted SVI per state. The latest SVI data available is from 2018, which means it was also calculated per 2010 census block groups. As specified earlier, the 2010 population data were used to calculate the weighted average to obtain SVI per state.

**Control Variables**

The control variables used in the statistical analysis of this study are at the state level and include the following: *population*, *the total number of tweets*, *number of unique users*, and *state race and ethnicity composition (%)*. For example, some states might have a higher population, and with that, it is very likely that more tweets will be collected in these states. These variables were included to avoid any bias that could appear due to the different numbers of tweets/sentiments in various states. State demographics include percentages of White, Black, Hispanic, and Asian populations per state, and they were also included to test if race and ethnicity play a role in the overall vaccination rate, HL, and SVI. Population and population percentages were calculated based on 2010 census group data since HL and SVI were also calculated per census 2010 data (2020 data was not available when HL and SVI were calculated).

**Statistical Analysis**

Various hypotheses were established to be tested by statistical analysis, such as:
**Hypothesis 1***:* Positive tweets are associated with a higher vaccination rate at the state level.
**Hypothesis 2**: Social determinants of health play a significant role in vaccine rate.
**Hypothesis 3**: A higher social vulnerability index is associated with lower health literacy.
**Hypothesis 4**: Percentages of minority populations are lower in states where health literacy is higher.
**Hypothesis 5**: Percentage of minority populations is higher in states where the social vulnerability index is more elevated.

The analysis consists of different statistical tests to evaluate each hypothesis. To test the first two hypotheses, a multivariant linear regression model was used to predict vaccination rate as a dependent numerical variable and tweet sentiments, health literacy, and social vulnerability index as independent variables. All the control variables mentioned were included in the models. Note that the tweets' sentiments were averaged per user per week, and state and week clustering were performed to avoid any data dependence due to the many values repetitions. For each user in the dataset, an average sentiment was calculated for that user per week. For that particular user the following data was available: location at the state level from which tweets were generated, HL & SVI for that state, vaccination rate per state for that particular week, that state race and ethnicity composition (percentages of White, Black, Hispanic and Asian population), population, the total number of tweets, and the number of unique users. The third hypothesis was also tested using multivariant linear regression, with SVI as dependent and HL as independent variables. Finally, the last two hypotheses were also tested using two separate linear regression models, one containing HL as dependent variable, and the second one containing SVI as dependent variable. In both models, independent variables were state race and ethnicity composition (%).

## Results

**Association Between Vaccination Rate and Tweet Sentiments and Impact of SDoH on Vaccination Rate**

The linear regression analysis showed a strong positive correlation between vaccination rate and tweets sentiments with the p-value < 0.05. This result supports **Hypothesis 1,** that positive tweets are associated with a higher vaccination rate. The results in **Table 2** show that both health literacy (HL) and social vulnerability index (SVI) play a significant role, with HL yielding a strong positive correlation while SVI is yielding a negative correlation. This means that for each unit increase of health literacy, the vaccination rate increases by 0.458 (p-value < 0.001). Similarly, for each SVI unit increase, the vaccination rate decreases by 14.1 (p-value < 0.001). These results support **Hypothesis 2,** that social determinants of health play a significant role in vaccination rates. Note that a similar analysis was performed using average tweet sentiments per user per day, and similar results were obtained.

**Table 2:** Association between vaccination rates and selected independent variables

| INDEPENDENT VARIABLES | ESTIMATE | STD. ERROR | T-VALUE | P-VALUE |
|---|---|---|---|---|
| Sentiment | 0.2734 | 0.1388 | 1.970 | P < 0.05 |
| Health Literacy | 0.4578 | 0.0586 | 7.812 | P < 0.001 |
| Social Vulnerability Index | -14.0801 | 1.2823 | -10.980 | P < 0.001 |
| White Population (%) | -9.4716 | 3.6687 | -2.582 | P < 0.05 |
| Black Population (%) | 3.1385 | 3.4969 | 0.897 | P > 0.05 |
| Hispanic Population (%) | 27.1188 | 1.6105 | 16.838 | P < 0.001 |
| Asian Population (%) | 31.0692 | 6.1772 | 5.030 | P < 0.001 |
| Population | 0.0000 | 0.0000 | 4.384 | P < 0.001 |
| Number of Unique Users | -0.0042 | 0.0003 | -16.480 | P < 0.001 |
| Total Number of Tweets | 0.0013 | 0.0001 | 20.105 | P < 0.001 |

**Association Between HL and SVI**

Another linear regression model was performed to test **Hypothesis 3**, having HL as explanatory variable and SVI as response variable. The results suggest that there is a significant negative correlation (P < 0.001) between HL and SVI. Thus, the social vulnerability index is lower in the states where health literacy is higher, suggesting that the areas with a higher ability to receive and understand health information are less vulnerable. This finding implies that public health interventions need to be performed in more vulnerable areas to educate communities and increase the vaccination rates. The relationship between HL and SVI is presented in **Figure 5**, confirming **Hypothesis 3**.

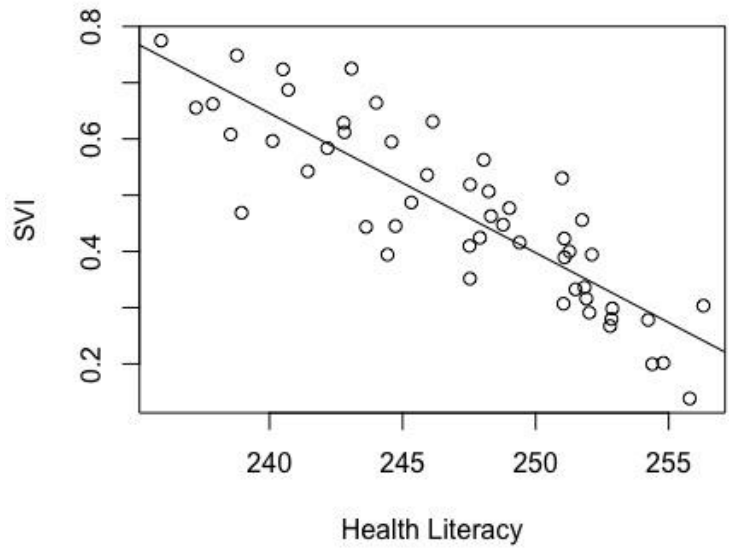

**Figure 5:** Linear relationship between SVI and HL

**Association Between HL/SVI and State Race and Ethnicity Composition**
Final linear regression models were used to test **Hypothesis 4** and **Hypothesis 5**, where dependent variables were HL in the first model, and SVI in the second model. The independent variables in both cases were state, race, and ethnicity composition (percentages of White, Black, Hispanic and Asian population). The result confirms **Hypothesis 4** by indicating the negative significant association between HL and percentages of Black and Hispanic population per state. Meaning, the HL is lower in the states with a higher percentage of Hispanic ($p < 0.001$) and Black ($p < 0.05$) population. Similarly, there is a negative significant relationship between SVI and White and Asian population percentage, and a positive significant correlation between SVI and Hispanic population percentage per state, confirming **Hypothesis 5**. This means that states with a higher social vulnerability index have a higher percentage of the Hispanic population ($p < 0.001$), while states with a lower SVI have a higher percentage of White ($p < 0.05$) and Asian ($p < 0.05$) population. This suggests that public health organizations need to dedicate resources to improve health education and increase vaccine awareness in vulnerable areas where there is a higher percentage of minority populations. Results of the models are presented in *Table 3.*

**Table 3:** Association between HL (above) and SVI (below) with state race and ethnicity composition

| INDEPENDENT VARIABLE (HL) | ESTIMATE | STD. ERROR | T-VALUE | P-VALUE |
|---|---|---|---|---|
| White Population (%) | 13.477 | 9.231 | 1.460 | P > 0.05 |
| Black Population (%) | -20.556 | 8.634 | - 2.381 | P < 0.05 |
| Hispanic Population (%) | - 31.423 | 3.673 | - 8.555 | P < 0.001 |
| Asian Population (%) | - 3.839 | 16.750 | - 0.229 | P > 0.05 |

| INDEPENDENT VARIABLE (SVI) | ESTIMATE | STD. ERROR | T-VALUE | P-VALUE |
|---|---|---|---|---|
| White Population (%) | - 0.9156 | 0.3830 | - 2.390 | $P < 0.05$ |
| Black Population (%) | - 0.0199 | 0.3583 | - 0.056 | $P > 0.05$ |
| Hispanic Population (%) | 0.8704 | 0.1524 | 5.711 | $P < 0.001$ |
| Asian Population (%) | - 1.5058 | 0.6950 | - 2.166 | $P < 0.05$ |

## Discussion

The purpose of this study was to examine the association between social media sentiments surrounding COVID-19 vaccination, the association between sentiments and vaccination rates in the US, and the importance of other social determinants of health factors that contribute to the COVID-19 vaccine hesitancy. The most important finding of this study from the public health perspective is the association between the state's vaccination rate and the tweets' sentiments. This indicates that social media could provide helpful information on vaccine acceptance, informing policymakers on what type of message would be beneficial for public health interventions on social media platforms. The use of social media by public health professionals could increase vaccine awareness and provide more detailed information about vaccination, contributing to the growth of vaccination rates. In addition, the analysis of social media activity could give early warnings about disease outbreaks, which public health organizations could use to prepare and guide their region/community-specific interventions proactively.

The potential future work would be identifying areas where social media sentiments are more negative than in others while analyzing the social vulnerability index, health literacy, demographics, and vaccination rates in these areas. Once those areas are determined, the next step would be to identify where health information can be distributed. One helpful tool to assist in this effort would be the Health Intelligence Atlas, a dashboard with multiple layers created by the research team that includes locations of public health agencies, libraries, places of worship, medically underserved areas and populations, etc.) [24]. Using a similar approach, it would be possible to identify vulnerable areas and perform public health interventions to increase the vaccination rates in these areas. This would be beneficial because the data found in this study can be added to this dashboard which might be used for future public health emergencies.

The study described herein offers a unique perspective using the social vulnerability index and health literacy to understand better the association between vaccination rate, social media sentiments, and these other potential factors that might impact them. The main strength of this study is that it was conducted on one of the most popular social media platforms, Twitter, a platform that can represent the public's opinions on vaccination. Another strength is that both tweets and vaccination rates data were collected daily in real-time, giving insight into trends surrounding vaccination and providing value to the analysis described in the study. Finally, all datasets were extracted from credible sources such as the CDC and the University of North Carolina. With the significant results from the analysis, this study is not without limitations: (1) the dataset consists of English tweets only, which might not reflect the vaccine hesitancy that is being discussed by minorities speaking a different language (e.g., Spanish, Vietnamese, etc.); (2) even though all the posts are related to the vaccination, some tweets might not represent an actual opinion of the user about it. Despite the limitations mentioned above, this study shows practical applications and results that are critical to future public health interventions.

## Conclusion

The primary purpose of this observational study was to investigate if any correlation exists between social media sentiments and the COVID-19 vaccination rates in the United States and identify other factors that impact the vaccination rate in the US. The regression analysis showed that the social media sentiments are significantly associated with the US COVID-19 vaccination rate. In addition, health literacy and social vulnerability index play an essential role in the COVID-19 vaccination rate. Social media might be used as an effective tool to increase the overall acceptance of public health interventions, such as the COVID-19 vaccination. Acknowledging that there are limitations, the results shown are relevant for future interventions and should be considered. Public health professionals should incorporate social media listening tools to analyze and address the spreading of negative posts about health interventions on social media and improve health literacy in socially vulnerable areas. As shown in this study, higher health literacy and more positive social media sentiments are significant factors in increasing the vaccination rate in the US. Therefore, social media is a reliable resource for informing the population about emerging threats/events and interventions, which is crucial in reducing misinformation and disinformation and building trust in the public health system.